\documentstyle[12pt]{article}

\begin{document}
\begin{flushright}
McGill/94-21\\ (May, 1994)\\
hep-th/9409083
\end{flushright}
\bigskip


\input epsf
\font\snall=cmr10
\font\bold=cmbx10


\begin{center}
{\large\bf STRING ORGANIZATION OF FIELD THEORIES:\\
DUALITY AND GAUGE INVARIANCE}\\

\bigskip\bigskip\bigskip
{Y.J. Feng$^{\dag}$ and C.S. Lam$^*$}\\
\bigskip
{Department of Physics, McGill University, 3600 University St.\\
Montreal, P.Q., Canada H3A 2T8}\\
\bigskip
\end{center}

\begin{abstract}
String theories should reduce to ordinary four-dimensional
field theories at low energies. Yet the formulation of the
two are so different that such a connection, if it exists, is not
immediately obvious. With the Schwinger proper-time representation,
and the spinor helicity technique, it has been shown that field
theories can indeed be written in a string-like manner, thus
resulting in simplifications in practical calculations, and
providing novel insights into gauge and gravitational theories.
This paper continues the study
of string organization of field theories by focusing on the question
of local duality.  It is shown that a single expression for the sum of
many diagrams can indeed be written for QED, thereby simulating the
duality property in strings. The relation between a single diagram and
the
dual sum is somewhat analogous to the relation between a  old-fashioned
perturbation diagram  and a  Feynman diagram. Dual expressions
are particularly significant for gauge theories because they are
gauge invariant while expressions for single diagrams are not.
\end{abstract}

\section{Introduction}

Superstring theories enjoy a number of interesting
properties at first sight not shared by ordinary
four-dimensional field theories.
 For a  superstring theory,
(1) the basic entity is
a string of size $O(10^{-32}$) cm, with massless levels and excitation
energies $O(10^{19}$) GeV;
(2) loop corrections are ultraviolet
finite and quantum gravity is well defined;
(3) the fundamental dynamical
variables consist of the spacetime
 $x^\mu(\sigma,\tau)$ and the
internal   $\psi^i(\sigma,\tau)$ fields, all as functions
of the worldsheet coordinates $\sigma$ and $\tau$; (4)
these variables propagate as independent
free fields throughout the worldsheet, in a
manner dependent on the topology but not on
the geometry of the worldsheet
(reparametrization and conformal invariance);
(5) an external photon of momentum
$p$ and wave function $\epsilon_\mu(p)$ is inserted into
the string through a vertex operator $\epsilon(p)\!\cdot\!
[\partial_\tau x (\sigma,\tau)]\exp[ip\!\cdot\!x(\sigma,\tau)]$, a form which
is
fixed by conformal invariance; (6) conformal invariance leads to local
(Veneziano) duality of the scattering amplitude.
A scattering process is described by
one or very few string diagrams. In particular, for elastic
scattering in the tree approximation, one string
diagram (the Veneziano amplitude)
gives rise simultaneously to all
 the $s$-channel and the
$u$-channel exchanges.

In constrast, in ordinary four-dimensional field theories (QFT),
($1'$) the basic entities are point particles;
($2'$) loop corrections
contain ultraviolet divergences and quantum gravity is
non-renormalizable;
($3'$) the fundamental dynamical
variables are  fields $\psi^i(x)$ of the four-dimensional
spacetime coordinates $x^\mu$; ($4'$) these fields propagate
freely only between vertices, where interactions take place; ($5'$)
external photons are inserted into  a Feynman diagram through the
photon operator $ \epsilon(p)\!\cdot\!A (x)\exp[ip\!\cdot\!x]$;
 ($6'$) there are
many distinct Feynman diagrams contributing to a scattering
amplitude. For elastic scattering, $s$- and $u$-channel exchanges
are given by different diagrams that must be added up together.

In the `low-energy' limit when $E\ll 10^{19}$ GeV, a string
of dimension $10^{-32}$ cm is indistinguishable from a point,
energy levels $O(10^{19}$) GeV are too high to matter,
so one would expect a string theory to reduce to an ordinary
field theory of zero masses.  In fact, explicit calculations
have been carried out \cite{1,2} to show that a one-loop $n$-gluon amplitude
in a string theory does reduce to corresponding results in field theory.
In order to go beyond one loop or the $n$-gluon amplitude, where no
simple string
expression is available, it is better to proceed in a
different way. Since the string properties (3)--(6)
are quite different from the field-theory properties ($3'$)--($6'$),
 the equivalence of string and QFT at low energies
is not immediately obvious. The
purpose  of this series of papers is, among other things, to make these
connections.

There are two motivations for doing so. At a theoretical
level, one can  hope to gain new insights into gauge and
gravitational theories from the string arrangement.
For example, according to (3) and ($3'$), spacetime and internal
coordinates
are  treated on an equal footing in a string, but not so
in QFT. This asymmetry  makes {\it local}
gauge transformations in QFT awkward to deal with and practical
calculations difficult to obtain. It would therefore be nice
to be able to reformulate gauge and a gravitational theories
in the symmetrical way of a string.
Conversely, one can hope that the successful multiloop treatment of
QFT, when written in a string-like manner,
 can give hints useful for   multiloop
string calculations.
On the practical side,
string-like organization of an field-theoretical amplitude  allows
 the spinor helicity technique \cite{3,4}
to be used on multiloop diagrams as easily as for tree diagrams \cite{5};
it also enables color, spin,
and momentum degrees of freedom to propagate independently, a separation
that
leads to efficient simplifications in actual calculations,
so much so that amplitudes not computable by ordinary means
 can be obtained
when organized in this novel way.
Examples of this includes the Parke-Taylor $n$-gluon amplitude \cite{6},
the factorization of  identical-helicity photons produced
in the $e^+e^-$ annihilation into $\mu $-pairs \cite{4},
the computation  of the one-loop $n$-photon amplitude
of identical helicities \cite{7}, and the calculation of the $n$-gluon
one-loop amplitudes \cite{1,2,8}.

In the low energy limit, the variable $\sigma$ along the string is
frozen, so the dynamical fields in (3) are effectively functions of
$\tau$ alone. To convert ($3'$) to (3), one must be able to free
the dependence of $\psi^i$ on $x$, to make   them both independent
functions of some {\it proper time} $\tau$. This can be accomplished by
using the Schwinger-parameter representation for the field-theoretic
scattering amplitude \cite{5}. In this representation, every vertex $i$ in
 a Feynman diagram
is assigned a proper time $\tau_i$, and each propagator $r$ is assigned
a
Schwinger parameter $\alpha_r$. If $r=(ij)$ is a line connecting
vertices $i$ to $j$, then $\alpha_r=|\tau_j-\tau_i|$.
If we regard the Feynman diagram as an electric circuit
with resistances $\alpha_r$, then  $x(\tau_i)$ can be interpreted as
(the four-dimensional) voltage at the vertex $i$ \cite{9}, thus freeing
it to be an independent function of $\tau$. Spacetime flow is thus
analogous to
the
current flow of an electric network, or equivalently the change of the
electrical potential from point to point.

Color flows can be isolated by creating color-oriented vertices in
Feynman
diagrams \cite{5}. One color-oriented vertex may be related to another by
twising, thus creating twisted Feynman diagrams in much the same
way like twisted open strings. The color subamplitudes so isolated
with these color-flow factors are gauge invariant.

With massless fermions,
chirality and helicities are conserved and this
allows free and unobstructed spin flows.
This is the essence of the spinor helicity technique which is applicable
to photons and gluons as well, for a spin-1 particle can kinematically
be regarded as the composite of two spin-${1\over 2}$ particles. The Schwinger
proper-time formalism allows the spinor helicity technique to continue
to work in loop diagrams \cite{5}; otherwise the loop momentum would be there
to
break chirality conservation and to obstruct the free flow of spins.

The spacetime, color, and spin flows thus obtained approximate the
properties
(3) and (4) of the string.
Moreover, using {\it differential circuit
identities} \cite{5}, external electromagnetic vertices can be converted
from ($5'$) to (5). What remains to be considered in completing the
string organization of field theory
is then the property of
{\it local duality}. Historically it was this unusual feature, first
seen in the Veneziano model \cite{10}, that marked the beginning of
string theory. To what extent this interesting property is
retained in field theory is therefore an interesting topic to study.
We propose to make a first attempt in that direction in the present
paper.

Duality for tree diagrams
is considered in Sec.~2, where its exact meaning
in field theory is also discussed.
For the sake of simplicity we shall confine that section
to a scalar theory where
a scalar photon field $A$ is coupled to a charged scalar meson field
$\phi$ via the Lagrangian $e\phi^*\phi A$, but it turns out
that once we solve the duality problem here it is solved in
quantum electrodynamics as well. To prepare for discussions of
QED and multiloop amplitudes, we review in Sec.~3 the
Schwinger-parameter representation for a field-theoretic scattering
amplitude which provides the main tool for further discussions.
Duality for the multiloop scalar theory is discussed in Sec.~4,
and duality for QED is discussed in Sec.~5. The problem of QCD
is much harder and we shall defer that to a future publication.
 Finally, a concluding
section appears in Sec.~6.

\section{Duality for Tree Diagrams in a Scalar-Photon Theory}

Duality in a string theory follows from its conformal invariance,
which in the special case of a four-point amplitude can be used to
fix the (complex) worldsheet positions of three of the external lines to
be
$0,1$ and $\infty$, and  the fourth one
 to be $x\in[0,1]$. A particularly
simple example is the Veneziano amplitude \cite{10}
in the Mandelstam variables $s$
and $u$,
\begin{eqnarray}
A(s,u)&&=-B(-u,-s)=-\int_0^1x^{-u-1}
(1-x)^{-s-1}dx \nonumber\\
&&=-{\Gamma(-s)\Gamma(-u)\over \Gamma(-s-u)}\ .
\end{eqnarray}
This amplitude has poles
when either $s$ or $u$ is a
non-negative integer (in units of $[M_P\sim O(10^{19})$ GeV]$^2$),
but there are no
simultaneous $s$- and $u$-channel poles. The amplitude can be
expanded either as a sum of $s$-channel poles,
represented purely by $s$-channel exchange Feynman diagrams, {\it or} a
sum
of $u$-channel poles,   represented  purely by $u$-channel
exchange diagrams.
There is no need to {\it add} both the $s$-channel and the
$u$-channel diagrams, as is necessary in ordinary quantum field
theories.
The $u$-channel amplitude is obviously equal to the $s$-channel
amplitude,
and both are equal to the single integral in (2.1).
This is {\it duality}.

These results can be obtained directly from
a pole expansion of the
Euler Gamma functions, or from the integral representation for the
Beta function.
In the latter case, an expansion of the integrand about $x=0$ gives rise
to the $u$-channel poles, and an expansion of the integrand about $x=1$
gives rise to the $s$-channel poles.

At present energies, $|s|, |u|\ll 1$ (in units of $M_P^2$), so only
the massless poles contribute, giving rise to
\begin{equation}
A(s,u)\simeq {1\over s} +{1\over u}\ .
\end{equation}
The $u$-channel pole comes from the divergence of the integral near
$x=0$
when $u=0$, and the $s$-channel pole comes from the divergence of the
integral near $x=1$ when $s=0$.
In this form, the amplitude does not {\it appear} to be `dual' anymore,
because
{\it both} the $s$-channel and the $u$-channel poles are {\it summed},
instead of having a {\it single} expression like (2.1), where only a sum
of the $s$-channel {\it or} a sum of the $u$-channel poles are present.
Nevertheless, appearances are deceiving, because (2.2) follows
mathematically from (2.1), which is dual.
In other words, the $u$-channel poles in (2.2)
can be formally obtained by an infinite sum of massive $s$-channel
poles,
and (2.2) is as dual as it can be at low energies.

It is instructive for later discussions to
 obtain (2.2) directly from the integral representation of (2.1).
For that purpose, divide the integral in $x$ into two
halves at the midpoint
$x={1\over 2}$. Since $|s|,|u|\ll 1$, the contribution of the integral comes
mainly from $x=0$, so we can put $(1-x)^{-s-1}\simeq 1$ there.
Similarly the
term $x^{-u-1}$ can be ignored in the second integral.
Next, make the transformation
$y=\ln(2x)$ in the first integral, so that $y\in[-\infty,0]$ there, and
the
transformation $y=-\ln[2(1-x)]$ in the second integral, so that
$y\in[0,\infty]$ there .
Then for $|s|,|u|\ll 1$,  the integral in (2.1) becomes
\begin{eqnarray}
A(s,u)&&\simeq-\int_{-\infty}^0  dy\exp(-uy)-\int_0^\infty dy\exp(sy)
\nonumber\\
&&={1\over u}+{1\over s}\ ,
\end{eqnarray}
which is the same as (2.2).

\begin{figure}
\vskip -2 cm
\centerline{\epsfxsize 4.7 truein \epsfbox {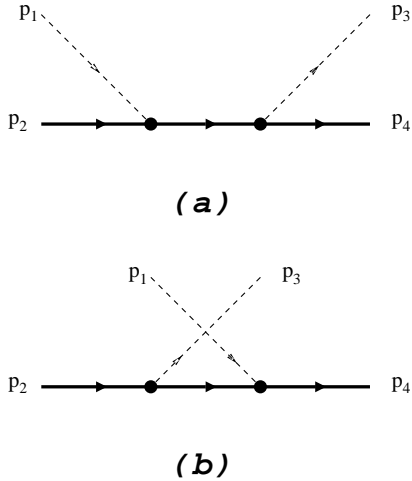}}
\nobreak
\vskip -7.5 cm\nobreak
\vskip .1 cm
\caption{Compton scattering diagrams,
in which $s=(p_1+p_2)^2$ and $u=(p_2-p_3)^2$}
\label{fig1}
\end{figure}

If one were to start directly from a massless scalar field theory
$\phi^*\phi
A$ (where all fields are scalar and massless), then the `Compton
scattering'
amplitude given by Fig.~1 is identical to (2.2). In that sense the
field-theoretic amplitude is already dual, or as dual as it can be
at the present energy range.
One might still be unsatisfied with this remark
about duality, and points out that
the original dual amplitude in (2.1) is given by a {\it single}
integral, whereas in (2.3) this is given by the sum of {\it two}
integrals.
Since (2.3) comes from (2.1) it must be able to write it as a single
integral as well. All that we have to do is to
define a function $P=\theta(-y)(-uy)+\theta(y)(sy)$, then
\begin{equation}
A(s,u)=-\int_{-\infty}^\infty dy\exp(P)\ ,
\end{equation}
which is of course equivalent to the Veneziano integral (2.1) at the
present energy range.  We
shall refer to expressions of this type, where
the sum of a {\it number of diagrams} is represented
by a {\it single} (possibly multi-dimensional) integral,
 as {\it dual expressions}.
What allows the dual expression (2.4) to be written is not so much the
explicit
form of the integrands in (2.3), but that the two integrals there have
 non-overlapping ranges in $y$. Given that, it is always possible to
define a
common integrand $\exp(P)$ so that the two integrals can be combined
into one.

Similar reasoning shows that dual expressions can be written for other
processes in a scalar-photon theory.
Since we are imitating the low energy limit of strings, we may simplify
writings by assuming all particles to be
 massless. In that case,
the Schwinger-parameter representation for a scalar propagator is
\begin{equation}
{1\over q^2+i\epsilon}=-i\int_0^\infty d\alpha\exp(i\alpha q^2)\ ,
\end{equation}
and the variable $\alpha$ is called a Schwinger proper-time parameter.
Using this, we can obtain (2.3) from (2.5) simply by letting
$i\alpha=-y$,  $q^2=u$ in the first term, and $i\alpha=y$,
$q^2=s$ in the second term. The fact that we
can get the Veneziano integral representation
(2.3) from the Schwinger representation
confirms the claim that the Schwinger-parameter representations
are string-like.

Consider now a more complicated example, Fig.~2, in
which charged particles scatter to produce $m$ (scalar) photons from one
charged line and $n$ photons from another.
We have drawn only one of the $(m+1)!(n+1)!$ possible diagrams; others
are obtained from it by permuting the photon lines.

\begin{figure}
\vskip .5 cm
\centerline{\epsfxsize 4.7 truein \epsfbox {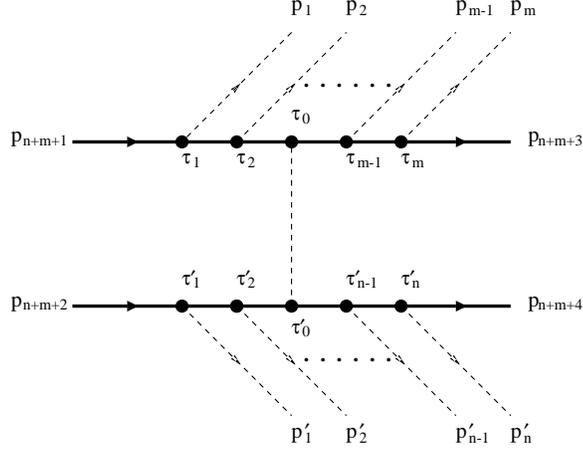}}
\nobreak
\vskip -10.5 cm\nobreak
\caption{A tree diagram for multiphoton
emission from charged-particle scattering}
\label{fig2}
\end{figure}

Let us first establish some common notations.
Assign to each vertex of a Feynman diagram a {\it proper time} $\tau$,
as illustrated in Fig.~2. The
Schwinger proper-time parameters $\alpha$ are then given by differences
of the proper times. Specifically,
if $r=(ij)$ is an internal line between vertices $i$ and $j$, then
$\alpha_r=|\tau_i-\tau_j|$. In this way,
all the proper-time differences are determined but translational
invariance
prevents the origin of proper time to be fixed. Let

\begin{eqnarray}
\int_a^b d\tau_{[12\cdots n]}&&\equiv
\int_a^bd\tau_1\int_{\tau_1}^bd\tau_2
\cdots\int_{\tau_{n-1}}^bd\tau_n\ ,\nonumber\\
\int_a^b d\tau_{[12(345)678]}&&\equiv \int_a^b d\tau_{[12]}
\int_{\tau_2}^bd\tau_{[678]}
\left(\prod_{i=3}^5\int_{\tau_2}^{\tau _6}d\tau_i\right)\ ,\nonumber\\
\langle\int d\tau_{[12\cdots n]}\rangle&&\equiv\lim_{T\to\infty}{1\over 2T}
\int_{-T}^{T}d\tau_{[12\cdots n]}\ ,\nonumber\\
\langle\int d\tau_{[12(345)678]}\rangle&&\equiv\lim_{T\to\infty}{1\over 2T}
\int_{-T}^T d\tau_{[12(345)678]}\ ,\nonumber\\
\langle\int d\tau_{(12\cdots n)}\rangle&&\equiv \langle\int d\tau_{[(12\cdots
n)]}\rangle
\ .
\end{eqnarray}
In short, the integration variable enclosed between square
brackets are ordered, and those between round brackets are unordered.

We can now return to Fig.~2. Its amplitude is
\begin{equation}
A=(-i)^{n+m+2}\langle\int d\tau_{[12\cdots 0\cdots m]}\rangle\langle\int
d\tau'_{[12\cdots 0\cdots n]}\rangle\exp(iP)\ ,
\end{equation}
for some quadratic function $P$ of the external momenta obtained by
using (2.5)
on each propagator. The detailed form of $P$ does not concern us at the
moment.
The other diagrams are obtained by permuting the photon lines
in Fig.~2, so their amplitudes are all given by something like (2.7),
but with
the $\tau$ and $\tau'$ integration regions permuted separately. The
detailed form of the quadratic function $P$ may change from region to
region
but again we do not have to worry about it.
Since the integration regions of these different diagrams do not
overlap,
it is possible to define a common $P(\tau,\tau')$
equal to the individual $P$'s
in their respective regions. In this way, all the $(m+1)!(n+1)!$
diagrams can be summed up to get a single dual expression
\begin{equation}
A_{sum}
=(-i)^{n+m+2}\langle\int d\tau_{(012\cdots m)}\rangle\langle\int
d\tau'_{(012\cdots
n)}
\rangle\exp(iP)\ ,
\end{equation}
in which the integration regions are completely unordered.

The reasoning can obviously be extended to any tree diagram in a
scalar-photon
theory. In every case, each charged line provides a platform for
ordering
the photon lines attached to it. Different diagrams correspond to
different
permutations of these photon lines, so they correspond to different
integration regions in the proper times. A sum into a single
dual expression in which the proper time integration regions are
unordered is
clearly possible. Furthermore, with minor modifications to be discussed
in Sec.~4, essentially the same consideration works
for multiloop diagrams as well.

There are three remarks to be made. First of all, it is
interesting to note that the relation between
a dual expression and a Feynman diagram
is very much like the relation between
a Feynman diagram and an old-fashioned diagram.
Recall that a Feynman diagram with $n$ vertices is made up
of a sum of $n!$ old-fashioned diagrams, corresponding to the $n!$
possible (real) time orderings of its vertices.
Similarly, a dual `diagram' consists of
a sum of Feynman diagrams, and they
differ from one another by the {\it proper-time} orderings of their
vertices.

Secondly, proper-time orderings in QED diagrams are very simple and
natural,
because there are the conserved charged lines
along which the photon vertices can be proper-time ordered. In contrast,
in a
neutral scalar $\phi^3$ theory, or in a pure gluon QCD, all lines
are equivalent and there are no obvious ways to proper-time order

\begin{figure}
\vskip -.5cm
\centerline{\epsfxsize 4.7 truein \epsfbox {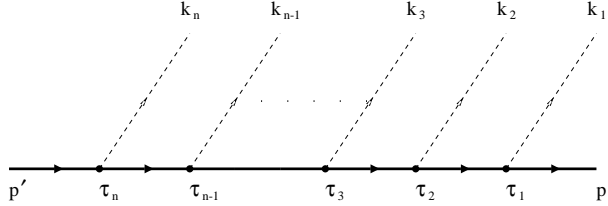}}
\nobreak
\vskip -13 cm\nobreak
\caption{ Multiphoton emission from a
charged line}
\vskip .1 cm
\end{figure}

\noindent the vertices,
especially in multiloop diagrams. This is one of the difficulties
one encounters in QCD.

Thirdly, there is the question of how useful these dual expressions are.
That naturally depends on the details of the diagrams one tries to sum,
and how simple the resulting integrand $\exp(iP)$ is. For example,
consider the emission of  scalar photons  shown in Fig.~3, where
$p$ and $k_i$ are massless but $p'$ may be offshell.
Using (2.5), and expressing the Schwinger parameters $\alpha $
as differences of the proper-time parameters $\tau _i$,
the exponent $P$ in the integrand $\exp(iP)$ becomes

\begin{eqnarray}
P&&=\sum_{i=1}^{n-1} \alpha _i(p+\sum_{j=1}^ik_i)^2 \nonumber\\
&&=2\sum_{i=1}^{n}(\tau _i-\tau _{n})p\!\cdot\!k_i
+2\sum_{i=1}^{n-1}\sum_{j=i+1}^{n}(\tau _j-\tau _{n})k_i\!\cdot\!k_j\ .
\end{eqnarray}
By using momentum conservation, this can also be written in
a more symmetric form:
\begin{eqnarray}
P=&&(\tau_1-\tau_{n})p'\!\cdot\!p+\sum_{i=1}^{n}(\tau_i-\tau_{n})p'\!\cdot\!k_i
- \nonumber\\
&&\sum_{i=1}^{n}(\tau_1-\tau_i)p\!\cdot\!k_i-{1\over 2}\sum_{i\not=j=1}^{n}
|\tau_i-\tau_j|k_i\!\cdot\!k_j\ .
\end{eqnarray}
The function $P$ for a permuted diagram can be obtained
from these expressions by permuting the photon momenta.

In general,
it is quite impossible to obtain a closed analytic expression for
its dual sum $A_{sum}$.
However, in the eikonal approximation where the
photon momenta are considered
small, $O(k_i\!\cdot\!k_j)$ terms can be neglected from (2.9), then
\begin{equation}
P\simeq  2\sum_{i=1}^n(\tau _i-\tau _{n})p\!\cdot\!k_i\ .
\end{equation}
The amplitude for Fig.~3 is then proportional to
\begin{equation}
A=(-i)^{n-1}\int_{\tau_n}^\infty d\tau_{[n-1,n-2,\cdots,2,1]}\exp(iP)\
,
\end{equation}
where $\tau_n$ is completely arbitrary. This freedom can be exploited
to render the amplitude
$A$ more symmetrical, if we multiply and divide it by
\begin{equation}
2p\!\cdot\!\left(\sum_{i=1}^nk_i\right)=-i\int_0^\infty
d\tau_n\exp\left(i\tau_n
2p\!\cdot\!\sum_{j=1}^n
k_j\right)\ .
\end{equation}
Then
\begin{equation}
A=2p\!\cdot\!\left(\sum_{i=1}^nk_i\right)(-i)^n\int_0^\infty
d\tau_{[n,n-1,\cdots,1]}\exp(i\tilde P)\ ,
\end{equation}
with
\begin{equation}
\tilde P=2\sum_{i=1}^n  \tau_ip\!\cdot\!k_i\ .
\end{equation}
Since $\tilde P$ is completely symmetrical in all the $k_i$, it is
identical
in all the permuted diagrams, so the dual sum of the $n!$
permuted diagram is
\begin{eqnarray}
A_{sum}&&=2p\!\cdot\!\left(\sum_{i=1}^nk_i\right)(-i)^n\int_0^\infty
d\tau_{(n,n-1,\cdots,1)}\exp(i\tilde P)\nonumber\\
&&=2p\!\cdot\!\left(\sum_{i=1}^nk_i\right)\prod_{j=1}^n{1\over 2
p\!\cdot\!k_j}\ ,
\end{eqnarray}
which is the well known eikenal expression.

Looking at this example, we see that there are two important ingredients
to make it successful.
The first is that $\tilde P$ has an identical functional form in every
one of the $n!$ integration regions. We shall refer to integrand
of this kind, that it has the same functional form in all integration
regions, to be {\it symmetrical}.
The second ingredient is that the
final integrals in (2.16) are simple enough to be computed analytically.
It turns out that the first ingredient is relatively easy to come by.
This is because the functional form of $P$ can be altered, either by
using momentum conservation to
substitute one external momentum by the negative sum of all others,
or by changing something like $\tau_i-\tau_j$ to $(\tau_i-\tau_k)+
(\tau_k-\tau_j)$. Quite often by making these changes one can manipulate
$P$ into a symmetric form. For example, eq.~(2.10) is completely
symmetrical in the indices $i=2$ to $n-1$ although (2.9)
is not.  Yet, without the second
ingredient, there is really not much point in achieving the first.
This is clearly seen by comparing (2.10) with (2.9). Without the
eikonal approximation, it is impossible to obtain $A_{sum}$
in either form, in spite of the symmetry of (2.10).
Even numerically it is not clear that
 a symmetric $P$ is easier
to compute, especially if its functional form is forced to
be very complicated when it is made symmetric.

\section{Schwinger-parameter Representation}

Every scattering amplitude can be written in the form
\begin{equation}
A=\left[{-i\mu^\epsilon\over
(2\pi )^{d}}\right]^{\ell }\int
\prod _{a=1}^{\ell }(d^{d}k_{a}){S_{0}(q,p)\over
\prod _{r=1}^{N}(-q_{r}^{2}+m_{r}^{2}-i\epsilon )}\ ,
\end{equation}
where $d=4-\epsilon$ is the dimension of spacetime,
$k_{a}\ (1\le a\le\l$) are the loop  momenta,
$q_{r}, m_{r}\ (1\le r\le N)$
are the momenta and masses of the internal lines, and $p_{i}\ (1\le i\le
n)$ are the outgoing external momenta.
Since we will be mainly interested in massless field theories, all
$m_r^2$
will be set equal to 0 in the following.

The numerator function
$S_{0}(q,p)$ contains everything except the denominators of the
propagators. Specifically, it is the product
of the vertex factors, numerators of propagators,
wave functions of the external lines,
symmetry factor, and the signs associated with closed fermion loops.
All the $i$'s and $(2\pi )$'s have been included in the factor before
the
integral.

By introducing a Schwinger proper-time
parameter $\alpha_{r}$ for each internal line to represent its scalar
propagator
as in (2.5),
the loop integrations in (3.1) can be
explicitly carried out to obtain the Schwinger-parameter representation
\cite{9}
\begin{equation}
A=\int [D \alpha ]\Delta (\alpha )^{-d/2}S(q,p)\exp[iP]\
,\end{equation}
where
\begin{eqnarray}
\int [D  \alpha ]\equiv&&\left[{(- i)^{d/2}\mu^\epsilon
\over  (4 \pi)^{d/2}}\right]^{\l }i^{N}
\int _{0}^{\infty }(\prod _{r=1}^{N}d\alpha _{r})\ ,\nonumber\\
P=&&\sum_{r=1}^N\alpha_rq_r^2\equiv\sum_{i,j=1}^nZ_{ij}(\alpha)p_i\!\cdot\!p_j\
,\end{eqnarray}
\begin{equation}
S(q,p)\equiv \sum _{k\ge 0}S_{k}(q,p)\ .\end{equation}
In spite of the same notation, the momenta $q_r$ in (3.2)
to (3.4) cannot be the same as the one in (3.1) since
loop-momentum integrations have now been carried out.  Instead, it is to
be
interpreted
as the current flowing through the $r$th line of an electric circuit
given
by the Feynman diagram, where $p_i$ are the outgoing currents  and
$\alpha_r$
are
the resistances of the $r$th line. With this interpretation, $P$ in
(3.2) and
(3.3) becomes the power consumed by the circuit, and $Z_{ij}$ is then
the
impedance
matrix. On account of current conservation, $\sum_{i=1}^np_i=0$, $q_r$,
$P$,
and hence the amplitude $A$
are invariant  under the {\it level transformation} $Z_{ij}\to
Z_{ij}+\xi_i+\xi_j$
for any arbitrary $\xi_i$. This enables us to choose an impedance matrix
with
$Z_{ii}=0$. Some of the formulas below, including (3.7), (3.11), and
(3.16),
are not valid without this condition.
Unless otherwise specified, this is
 a choice we will adopt throughout.
Note that these and the following formulas are equally valid for tree
diagrams, or a combination of trees and loops.

Suppose the numerator function $S_{0}(q,p)$ is a
polynomial in $q$ of degree
$j$. Then $S_{k}(q,p)$ is defined to be
 a polynomial in $q$ of degree $j-2k$, obtained from $S_{0}(q,p)$ by
 contracting $k$ pairs of $q$'s in all possible ways  and
 summing over all the contracted results. The rule for contracting a
pair of $q$'s is:
\begin{equation}
q^{\mu }_{r}q^{\nu }_{s}\to -{i\over 2}H_{rs}(\alpha)g^{\mu \nu }
\equiv q^{\mu }_{r}\sqcup q^{\nu }_{s}\ .\end{equation}
The circuit quantities in (3.2) to (3.5), including
$\Delta,\ q_r,\ P,\ Z_{ij}$ and $ H_{rs}$, can all be
obtained directly from the Feynman diagram \cite{9}. For example,
the formula for the impedance matrix and the function $\Delta$ are
\begin{equation}
\Delta=\sum_{T_1}\prod^\l \alpha \ ,
\end{equation}
\begin{equation}
Z_{ij}=-{1\over 2}\Delta^{-1}\sum_{T_2^{ij}}\prod^{\l+1}\alpha\ .\end{equation}
These formulas have the following meaning. An $\l$-loop diagram can
be turned into a tree by cutting $\l$ appropriate lines.
 $\Delta$ in (3.6) is obtained by summing over the set $T_1$
of all such cuts, with
the summand being the product of the $\alpha $'s of the cut lines
in each case. For tree diagrams where$\l=0$, by definition $\Delta=1$.
 Similarly, let $T_2^{(ij)}$ be the set of all cuts
of $\l+1$ lines so that the diagram turns into two disconnected
trees, with vertex $i$ in one tree and vertex $j$ in another.
Then $-2\Delta\!\cdot\!Z_{ij}$ in (3.7) is given by the sum over $T_2^{(ij)}$,
with
the summand being the product of $\alpha $'s of the cut lines.

Besides satifying Kirchhoff's law, the electric-circuit
quantities obey a number of {\it differential circuit identities} \cite{5}:
\begin{equation}
{\partial \over\partial \alpha _{r}}P(\alpha ,p)=q_{r}^{2}\ ,\end{equation}
\begin{equation}
{\partial \over\partial \alpha _{s}}q_{r}(\alpha ,p)
=H_{rs}q_{s}\ ,\end{equation}
\begin{equation}
{\partial \over\partial \alpha
_{t}}H_{rs}(\alpha)=H_{rt}(\alpha)H_{ts}(\alpha)\
.\end{equation}
Moreover, the {\it contraction function} $H_{rs}=H_{sr}$
is `conserved' at each vertex as if the external
currents were absent, {\it i.e.,}
if $\sum_{r\in
V}q_r=p$ is obeyed at some vertex, then $\sum_{r\in V}H_{rs}=0$ for all
$s$.
In particular, if $q_r$ does not involve $\alpha$'s as is the case when
it
is a branch of a tree,  then $H_{rs}=0$ for all $s$.

So far all quantities are expressed as functions of $p_i$ and
$\alpha_r$.
As in the case for tree diagrams, we can assign each vertex with a
proper time $\tau_i$ and consider $\alpha_r=|\tau_i-\tau_j|$ if
$r=(ij)$.
We can then convert all $\alpha$-integrations into $\tau$-integrations.

\begin{figure}
\vskip -0 cm
\centerline{\epsfxsize 4.7 truein \epsfbox {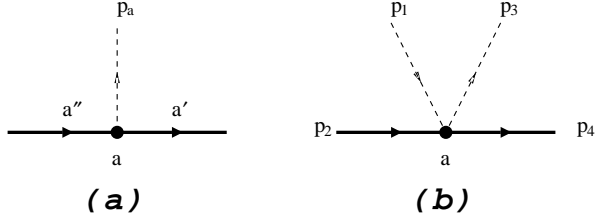}}
\nobreak
\vskip -13 cm\nobreak
\caption{Cubic vertex $C_a$ and seagull
vertex $Q_a$ of scalar electrodynamics}
\vskip .1 cm
\end{figure}

In scalar QED, there are two kinds of vertices: the cubic vertex
$C_a=e \epsilon (p_a)\!\cdot\!(q_{a'}+q_{a''})$ in Fig.~4(a) and the seagull
vertex
$Q_a=2e^2\epsilon(p_1)\!\cdot\!\epsilon(p_3)$ in Fig.~4(b). We shall call a
cubic
vertex `external', and perhaps less confusingly of {\it type-a}, if
it consists of one external photon line and two internal charged-scalar
lines. A type-a vertex $a$ has a string-like representation \cite{5}
\begin{equation}
C_a=-ie \epsilon (p_a)\!\cdot\!D_a(iP)\ ,\end{equation}
where
\begin{equation}
D_aP\equiv \partial_a{\partial P\over \partial p_a}\ , \quad
\partial_a\equiv{\partial\over\partial
\tau_a}\ .\end{equation}
Unfortunately, the same representation is not true for internal
cubic vertices.

To make (3.12) useful, we shall define $D_a$ to operate on
functions of the form $f(\tau)\!\cdot\!$ $\prod_rq_r\exp(iP)$ like a
derivation,
{\i.e.,} like a first derivative satisfying the product rules but
not like a second derivative. To complete the definition, we must also
define $D_a$ when it operates on $f(\tau ), P$ and $q_r$. For each of
these three elementary operations, it will be $D_a=\partial_a(\partial/\partial
p_a)$
as in (3.12). This leads to $D_a f(\tau )=0$ and it can be shown that
\cite{5}
\begin{equation}
D_aq_r=H_{ar}\ , \quad D_a^\mu D_b^\nu P
=2g^{\mu \nu }H_{a'b'}\ ,\end{equation}
where $b$ is another type-a vertex, and  $a'\not= r\not= a''$.

Eq.~(3.11) makes it possible to replace a vertex $C_a$ by an operation
involving $D_a$. Eq.~(3.13) shows that the necessary contractions
(3.5) can automatically be accommodated as well.
Consequently, a scalar QED amplitude with $n_a$ vertices of type-a
can be written as \cite{5}
\begin{eqnarray}
A=&&\int [D\alpha]\Delta^{-d/2}(-ie)^{n_a}[\epsilon (p_1)\!\cdot\!D_1]
[\epsilon (p_2)\!\cdot\!D_2]\cdots\nonumber\\
&&[\epsilon (p_{n_a})\!\cdot\!D_{n_a}]S^{int}(q,p)\exp(iP),\end{eqnarray}
where $S^{int}(q,p)$ is given by the product of the non-type-a
 vertices and terms generated from their mutual contractions.
Moreover, it is true that \cite{5}
\begin{equation}
\partial_a\Delta=0
\end{equation}
for a type-a vertex,
so it does not matter whether $\Delta$ in (3.14) is put before or after
the $D_a$'s. Another useful relation to know is
\begin{equation}
\partial_a Z_{ij}=0\end{equation}
provided $i\not=a\not=j$. This relation can be used to show how the
string-like vertex changes under a guage transformation,
when $\epsilon (p_a)$ is replaced by $p_a$. In that case, remembering
that $Z_{aa}=0$, (3.3) and (3.16) give
\begin{eqnarray}
C_a&\to -ie\partial_a\left(p_a\!\cdot\!{\partial P\over\partial p_a}\right)=
-2ie\partial_a\left(p_a\!\cdot\!\sum_iZ_{ai}p_i\right)&\nonumber\\
&=-ie\partial_a\sum_{i,j}Z_{ij}p_i\!\cdot\!p_j=-ie\partial_aP\ .&
\end{eqnarray}

\section{Multiloop Duality for the Scalar Theory}

In the scalar-photon theory with interaction $\phi ^* \phi  A$, the most
general amplitude is given by (3.2), with $S=1$ and $P$
given by (3.3) and (3.7). The Schwinger parameters $\alpha $
will be expressed as differences of the proper-time parameters $\tau $,
and the proper times will again be ordered along the charged
lines as in the tree cases. The only new problem here is where to
begin the ordering in the case of a charged-scalar loop.
\begin{figure}
\vskip -0 cm
\centerline{\epsfxsize 3.0 truein \epsfbox {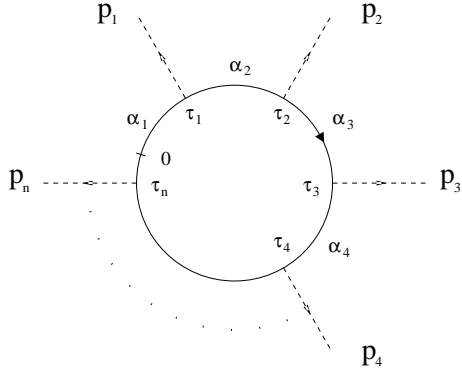}}
\nobreak
\vskip -5 cm\nobreak
\caption{An $n$-photon one-loop diagram}
\end{figure}

Since the origin of the proper time is never determined by the
$\alpha $'s, it can be chosen arbitrarily, say at the position
marked `0' in Figs.~5 and 6. We must now insert into (3.2) a factor
\begin{equation}
1=\int_{0}^\infty dT \delta (\sum_{loop} \alpha -T)\end{equation}
for every charged loop, where the sum is taken over
all the $\alpha $'s in the loop.
So for Fig.~5, the $\alpha $-integrations can be replaced by
\begin{equation}
\prod_{i=1}^n\left(\int_0^\infty d\alpha _i\right)=\int_0^\infty {dT\over T}
\int_0^T d \tau _{[12\cdots n]}\equiv
\langle\int d \tau _{[12\cdots n]}\rangle\ ,\end{equation}

\begin{figure}
\vskip -0 cm
\centerline{\epsfxsize 4.7 truein \epsfbox {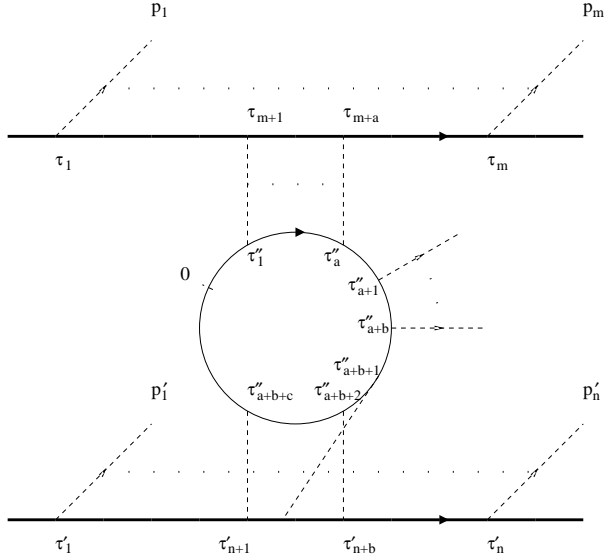}}
\nobreak
\vskip -9.5 cm\nobreak
\caption{A complicated multiloop scattering
diagram}
\vskip .1 cm
\end{figure}

\noindent
where $\alpha _i=\tau _{i+1}-\tau _i$ for $i\le n-1$, and
$\tau _n=T-(\tau _n-\tau _1)$.
Strictly speaking, there is an inconsistency in the definition
of $\langle d \tau _{[\cdots]}\rangle$  between (4.2)
and (2.6), but
in practice (4.2) is always used for closed charged loops and (2.6)
is always used for open charged lines.

One can now obtain  dual amplitudes by summing over all photon
permutations
in exactly the same way as before.
For example, for Fig.~5 and its permuted diagrams, the sum is
proportional to
\begin{equation}
A_{sum}=\langle\int d \tau _{(12\cdots n)}\rangle
\Delta^{-d/2}\exp(iP)\ .\end{equation}
For Fig.~6 and its permuted diagrams, the sum is proportional to
\begin{eqnarray}
A_{sum}=&&\langle\int d \tau _{(12\cdots m+a)}\rangle
\langle\int d \tau' _{(12\cdots n+b)}\rangle\langle\int d \tau'' _{(12\cdots
a+b+c)}\rangle\nonumber\\
&&\Delta^{-d/2}\exp(iP)
\ .\end{eqnarray}

As in tree amplitudes, how useful such  dual expressions are depends
on the complexity of $P$ and $\Delta$ in each case. In the
eikonal approximation (2.11)---(2.16), the integrals can be
carried out because $\tilde P$ is common for all integration regions
{\it and}
because it has a simple dependence on the proper times. Now
the first condition is not hard to meet, if the diagrams to be summed
have a high degree of symmetry. For example, for the well-studied
case of Fig.~5 \cite{8,11,5}, eqs.~(3.6) and (3.7) show that
$\Delta=\sum_{i=1}^n
\alpha _i=T$, and
\begin{equation}
Z_{ij}=-|\tau _i-\tau _j|\left(T-|\tau _i-\tau _j|\right)/2T
\equiv G_B(\tau _i,\tau _j)
\ ,\end{equation}
which has a symmetric form in the sense that it has the same functional
form in
all integration regions. However, one is still unable to evaluate the
integral (4.3) analytically because of its relatively complicated
$\tau$-dependences.

For the reason discussed at the end of Sec.~2, $P$ can be made symmetric
or partially symmetric in many cases.
One way of seeing this is the following.
If we modify Fig.~3 by adding on a charged-scalar
propagator at each end, then analogous to (2.10) one can produce a form
of $P$ which is completely symmetrical in all the
photon lines. We have already seen that the $P$ in Fig.~5 is completely
symmetrical in all its photon lines as well.  Now every Feynman diagram
can be built up from a number of open charged lines with
its attached photons, and a number of one-charged-loop diagrams with
its attached photons, by joining together pairs of photon lines.
Mathematically, one obtains the resulting amplitudes by multiplying
these $P$'s of the components,
together with the propagators of the joined photon lines
in the form of (2.5), then carries out the momentum integrations
of the joined photon lines. Since the dependences on these joined
momenta are
Gaussian, such integrations can be carried out, and one again obtains
a result of the form (3.2), with $S(q,p)=1$, and with $P$ of (3.2)
a function of the component $P$'s. Since the component $P$'s are
symmetric in all the photon lines, they will still be symmetric
in the remaining, unjoined, external photon lines, so in this way
one can obtain a symmetric form for the final
 $P$. This mechanism for obtaining
a symmetric form has
been discussed recently \cite{12} in a slightly different language.
However, the symmetric form obtained this way is usually much more
complicated than those obtained directly from (3.3), (3.6), and (3.7).
It is so complicated that it is unlikely to be integrated analytically,
nor will it lead to simpler numerical evaluations in most cases.
See the end of Sec.~2 for more discussions along these lines.
Though one can generally produce other simpler symmetric forms, they are
still not simple enough for the integrations to be carried out
explicitly.
For these reasons there seems to be no particular advantage of having
a symmetric form and we will not do so most of the time.

\section{QED}

The most important ingredient for obtaining a dual expression
is the presence of a conserved charged line along which to order
the interaction vertices. This is independent
of the spin of the particles involved, hence one can
obtain dual amplitudes in QED just as easily as those in the
scalar-photon
theory.

The factor $S(q,p)$ in (3.2) is no longer 1 for QED. For scalar
QED, for example, it is made up of the product of the vertex factors
$C_i$ and $Q_i$ and their contractions. See eq.~(3.11) and the
paragraph above it. This however makes no difference to the
construction of the dual amplitude. Another minor complication
is the presence of the seagull vertex $Q_i$. This simply adds
a Dirac-$\delta $ function contribution to the integrand.
For example, in the Compton scattering diagram Fig.~7,
all that we have to do to accommodate the seagull vertex in Fig.~7(c)
is to define the integrand of the dual amplitude to be
\begin{eqnarray}
S\exp(iP)&&=\theta (\tau _3-\tau _1)S_a\exp(iP_a)
+\theta (\tau _1-\tau _3)S_b\exp(iP_b)\nonumber\\
&&+\delta (\tau _3-\tau
_1)S_c\exp(iP_c)
\ ,
\end{eqnarray}
where $S_i,P_i\ (i=a,b,c)$ are the respectively factors for the
three diagrams Figs.~7(a), 7(b), and 7(c).
\begin{figure}
\vskip -2.5 cm
\centerline{\epsfxsize 4.7 truein \epsfbox {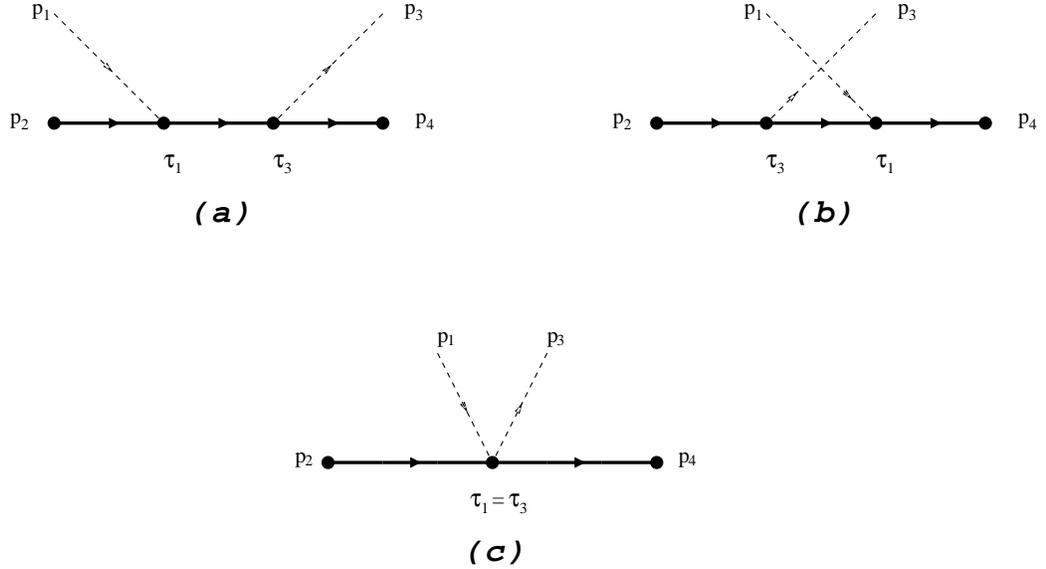}}
\nobreak
\vskip -5.5 cm\nobreak
\caption{Lowest order Compton scattering
diagrams in scalar QED}
\vskip .1 cm
\end{figure}

What distinguishes the dual expressions of QED from the
 scalar theory is gauge invariance, in that   the diagrams in QED
to be summed are connected by gauge invariance.
This means that the gauge-dependent parts of each diagram should
no longer be there in the dual sum. How the dual expression can
be mathematically manipulated to achieve this purpose unfortunately
depends on the details. At this moment two general techniques are
available to aid us. One is the spinor helicity technique \cite{3,4,5}, where
the reference momenta of the photons can be chosen appropriately
to reduce the amount of gauge-dependent contributions. The other
is the integration-by-parts technique \cite{1,2,11} used in connection
with the string-like operators appearing in (3.11) and (3.14).

We mentioned previously that the relation between a dual amplitude
and a Feynman amplitude is analogous
to the relation between
a Feynman amplitude and an old-fashioned amplitude. In each case
the former is not time-ordered, and the latter is; the only difference
being that it is  proper-time ordering for the first pair and real-time
ordering for the second. Now with gauge theories, there is another
parallel  between these two cases: an old-fashioned diagram
is not relativistically invariant but a Feynman diagram is. Similarly,
a Feynman diagram is not gauge invariant but a dual diagram is.

Let us consider two simple examples to illustrate these points.
First, consider the Compton amplitude Fig.~7 in scalar QED.
A propagator is added to each external charged line so that
the vertices 1 and 3 in Figs.~7(a) and 7(b) are of type-a,
to enable the string-like vertex (3.11) to be used.
Without the seagull term Fig.~7(c), the amplitude is not gauge
invariant, so it definitely contains a non-trivial gauge-dependent
part. As we shall see below, one can actually manipulate the dual
expression so that the seagull vertex seems to disappear,
and the gauge-dependent parts from these three diagrams are no longer
present.

The expression $P$ for diagrams (a), (b), and (c) are respectively
\begin{eqnarray}
P_a&&=(\tau_3-\tau_1)(p_1+p_2)\!\cdot\!(p_3+p_4)
+(\tau_1-\tau_2)p_2\!\cdot\!(p_3+p_4-p_1)\nonumber\\
&&+(\tau_4-\tau_3)p_4\!\cdot\!(p_1+p_2-p_3)\ ,\nonumber\\
P_b&&=(\tau_1-\tau_3)(p_2-p_3)\!\cdot\!(p_4-p_1)
+(\tau_3-\tau_2)p_2\!\cdot\!(p_3+p_4-p_1)\nonumber\\
&&+(\tau_4-\tau_1)p_4\!\cdot\!(p_1+p_2-p_3)\ ,\nonumber\\
P_c&&=\left(P_a\right)_{\tau _3=\tau _1}=\left(P_b\right)_{\tau _3=\tau _1}
\ .\end{eqnarray}
Using (3.14), the vertex factors in both cases become
\begin{equation}
S_a=S_b=(-ie)^2\left[\epsilon (p_1)\!\cdot\!D_1\right]\left[\epsilon
(p_3)\!\cdot\!D_3\right]
\equiv S\ .\end{equation}
In their respective regions, one can replace $P_a$ and $P_b$ by
\begin{eqnarray}
P'_a&=\theta(\tau _3-\tau _1)P_a&\nonumber\\
P'_b&=\theta(\tau _1-\tau _3)P_b
\ .&\end{eqnarray}
However, since $\tau $-differentiations are involved in $S$ of (5.3),
$S_a\exp(iP'_a)$ and $S_b\exp(iP'_b)$ are not identical to
$S_a\exp(iP_a)$ and $S_b\exp(iP_b)$. It can be checked by explict
calculation that the former already contains the seagull vertex
in Fig.~7(c). Hence the dual expression for Fig.~7 is
\begin{equation}
A_{sum}=-ie^2\langle\int_{\tau _2}^{\tau _4}d \tau _{(13)}\rangle
 \left[\epsilon(p_1)\!\cdot\!D_1\right]\left[ \epsilon(p_3)\!\cdot\!D_3
\right]\exp(iP')\ ,\end{equation}
where
\begin{equation}
P'=\theta(\tau_3-\tau_1)P_a+\theta(\tau_1-\tau_3)P_b=P'_a+P'_b\
.\end{equation}
The fact that the seagull vertex {\it seems} to have disappeared
suggests that we have eliminated the gauge-dependent contributions
altogether. To see that explicitly, use (3.17). Then under a gauge
transformation, $\epsilon (p_a)\!\cdot\!D_a$  ($a=1,3$) is changed into
something
proportional to $\partial_a$. The integral over $\tau _a$ in (5.5)
can then be carried out to yield the boundary contributions
at $\tau _2$ and $\tau _4$, and hence the Ward-Takahashi identity.
Since there is no trace of explicit cancellations needed
at $\tau _1=\tau _3$, it must mean that the gauge-dependent
terms in the individual diagrams have now been eliminated.

Another simple example one can mention is QED in the eikonal
approximation, Fig.~3. In the soft photon limit for scalar
QED, diagrams involving seagull vertex are not dominant
because of the presence of one less propagator $O(k^{-1})$
in the amplitude. The cubic vertex factor is trivial in the
soft photon limit, yielding $S=\prod_{i=1}^n(2e \epsilon (k_i)\!\cdot\!p)$.
The dual amplitude can therefore be read off from (2.17) to be
\begin{equation}
A_{sum}=2p\!\cdot\!\left(\sum_{i=1}^nk_i\right)\prod_{j=1}^n
\left({e \epsilon (k_j)\!\cdot\!p\over  p\!\cdot\!k_j}\right)\
\end{equation}
It is gauge invariant to leading order in the photon momenta.

\section{Conclusion}

In the Schwinger-parameter representation, QED diagrams differing
from one another by the permutation of photon lines correspond to
different proper-time ordering of the vertices, and can be formally
summed into a single integral over a hypercubic region.
This sum is referred to as a dual sum because it is  the field-theoretic
counter part of a dual amplitude in string theory.
The relation between individual Feynman diagrams and their
dual sum is analogous to the relation between individual old-fashioned
diagrams and their sum into a single Feynman diagram. Among other
things,
individual Feynman diagrams in QED are not gauge invariant but the
dual sum is. Similarly, the individual old-fashioned diagrams are not
Lorentz invariant but their sum is. The dual sum allows formal
manipulations
between different Feynman diagrams to be carried out, {\it e.g.,} by
the integration-by-parts technique on string-like vertices. With
appropriate approximations, such as the eikonal approximation, explicit
results may sometimes be obtained from the dual expression as well.

Dual expressions for QCD are much more complicated to deal with and are
not
discussed in the present paper.

\section{acknowledgement}
This research is supported in part by the Natural
Sciences and Engineering Research Council of Canada and the Qu\'ebec
Department of Education.s


\begin{thebibliography}{99}

\bibitem[\dag]{YF}e-mail Address: feng@physics.mcgill.ca.
\bibitem[*]{CL}e-mail Address: lam@physics.mcgill.ca.
\bibitem[1] 1  Z. Bern and D.K. Kosower, { Phys.~Rev.~Lett. }{\bf B66} (1991),
1669; { Nucl.~Phys. }{\bf B379} (1992), 451.
\bibitem[2] 2 Z. Bern and D.C. Dunbar, { Nucl.~Phys. }{\bf B379} (1992), 562.
\bibitem[3] 3 P. De Causmaecker, R. Gastmans,
W. Troost, and T.T. Wu, { Phys.~Lett. }{\bf 105B} (1981), 215;   { Nucl.~Phys.
}{\bf B291} (1987), 392;
 M.L. Mangano and S.J. Parke, { Phys. Rep.}  200 (1991), 301;
R. Gastmans and T.T. Wu, `The Ubiquitous Photon',
International Series of Monographs on
Physics, Vol.~80 (Clarendon Press, Oxford, 1990).
\bibitem[4] 4 Z. Xu, D.-H. Zhang,
and L. Chang, Tsinghua University Preprints,
Beijing, China, TUTP-84/4, TUTP-84/5, TUTP-84/6.
\bibitem[5] 5 C.S. Lam,  { Nucl.~Phys. }{\bf B397} (1993), 143; { Phys.~Rev.
}{\bf D48} (1993), 873;
{ Can. J. Phys. } to appear.
\bibitem[6] 6 S. Parke and T. Taylor, { Phys. Rev. Lett.} 56 (1986),
2459.
\bibitem[7] 7 G. Mahlon, Fermilab preprint Fermilab-Pub-93/327-T (1993).
\bibitem[8] 8 Z. Bern, L. Dixon, and D.A. Kosower,
{ Phys.~Rev.~Lett. }{\bf B70} (1993), 2677;
Z. Bern, L Dixon, D. C. Dunbar, and D. A.
Kosower,  SLAC preprint SLAC-PUB-6415.
\bibitem[9] 9 C.S. Lam and J.P. Lebrun, { Nuovo Cimento} {\bf 59A} (1969),
397.
\bibitem[10] {10} G. Veneziano, { Nuovo Cimento} {\bf 57A} (1968), 190.
\bibitem[11] {11} M. Strassler, { Nucl.~Phys. }{\bf B385} (1992) 145; SLAC
preprint SLAC-PUB 5978 (1992).
\bibitem[12] {12} M.G. Schmidt and C. Schubert, DESY preprint
DESY-HD-THEP-94-7 (hep-ph/9403158) (1994).
\end{thebibliography}
\end{document}